# Tutorial: Exciton resonances for atomically-thin optics

Jason Lynch[1], Ludovica Guarneri[2], Deep Jariwala[1,*] and Jorik van de Groep[2,*]

[1] Electrical and Systems Engineering, University of Pennsylvania, Philadelphia, PA, USA
[2] Van der Waals-Zeeman Institute, Institute of Physics, University of Amsterdam, Amsterdam, The Netherlands

*Corresponding author email(s): j.vandegroep@uva.nl; dmj@seas.upenn.edu

**Abstract**

Metasurfaces enable flat optical elements by leveraging optical resonances in metallic or dielectric nanoparticles to obtain accurate control over the amplitude and phase of the scattered light. While highly efficient, these resonances are static and difficult to tune actively. Exciton resonances in atomically-thin 2D semiconductors provide a novel and uniquely strong resonant light-matter interaction, which presents a new opportunity for optical metasurfaces. Their resonant properties are intrinsic to the band structure of the material and do not rely on nanoscale patterns and are highly tunable using external stimuli. In this tutorial, we present the role that excitons resonances can play for atomically-thin optics. We describe the essentials of metasurface physics, provide a background on exciton physics, as well as a comprehensive overview of excitonic materials. Excitons demonstrate to provide new degrees of freedom and enhanced light-matter interactions in hybrid metasurfaces through coupling with metallic and dielectric metasurfaces. Using the high sensitivity of excitons to the medium's electron density, the first demonstrations of electrically-tunable nanophotonic devices and atomically-thin optical elements are also discussed. The future of excitons in metasurfaces looks promising, while the main challenge lies in large-area growth and precise integration of high-quality materials.

## 1. Introduction

The research field of flat optics and photonics aims to realize evermore compact and functional optical elements by manipulating the flow of light at the nanoscale. Based on improved understanding into light scattering by nanostructures, bulky optical elements are now being replaced by compact arrays of carefully designed nanostructures. Light scattering by nanostructures enables accurate control over the flow of light. Electromagnetic energy from the incident wave couples to the resonant modes in the nanostructure and the subsequent re-radiation of light by the oscillating charges inside the nanostructure gives rise to the scattered fields. By controlling the nanostructure material, geometry, and dielectric surrounding, the amplitude and phase of the scattered fields can be engineered at will. Metal nanostructures were already studied by Michael Faraday in 1857[1], who characterized light scattering by gold nanoparticles in water and glass. Since then, multiple seminal studies have contributed to the (analytical) description of light scattering, including Mie theory in 1908[2]. Over the past decades in particular, the field of nanophotonics has developed a deep understanding of such resonant light scattering processes by both metal and dielectric nanostructures[3,4].

Inspired by Huygen's principle, nanostructures can be placed in sub-wavelength arrays – referred to as *metasurfaces* – to function as a 2D surface covered with physical point sources of scattered light. The interference of these scattered fields forms a new wave front and can thus be directed at will through accurate control over the relative phase difference between the nanostructures. As such, optical resonances lie at the heart of metasurface physics, and the optical functionality of a metasurface can be controlled by engineering the optical resonances in the individual nanostructures. Metasurface flat optics now form the basis for a rapidly emerging research field. Based on advanced design concepts such as dispersion engineering[5], combining degenerate resonances[6,7], and computational inverse design techniques[8,9], nanophotonic metasurfaces are now being explored for a wide range of applications, including high-numerical aperture lensing[10,11], 3D multi-color holography[12–15], and beam steering[16–18]. The current state of the field of metasurfaces is the topic of multiple excellent review papers[19–21].

Despite these rapid advances, metasurface functions have largely remained static with their functionality fixed at the moment of nanofabrication. At the same time, newly emerging and future application of metasurfaces such as light detection and ranging (LIDAR), computational imaging, augmented/virtual reality, and sensing, require dynamic control of the metasurface function[22,23]. Current approaches to manipulate optical resonances include phase-change materials[24–26], mechanical movement[27–29], and electrostatic gating[30–34] amongst others. However, these approaches are typically slow, require complex 3D architectures, or are limited to the infrared spectral range. The applicability of plasmonic and Mie resonances in dynamic metasurfaces has been limited due to the small magnitude of most electrorefraction and electroabsorption effects in metals and semiconductors[35] – i.e. the complex refractive index shows only small changes in the presence of excess charge carriers or electrical fields. It is therefore clear that the next-generation metasurfaces demand novel materials that exhibit a strong and tunable light-matter interaction.

In parallel, the past decade has experienced an unprecedented interest in the study of 2D layered materials. Soon after the discovery of graphene[36], 2D materials from a wide range of electronic material classes (metallic, insulating, semiconducting) were identified with intriguing electronic as well as optical properties. These so-called van-der-Waals (vdW) materials (named after the van-der-Waals force that bonds the 2D atomic layers) can be exfoliated down to the monolayer limit[37] and combined into heterostructures without the need for lattice matching[38] through e.g. deterministic stamping methods[39].

In the monolayer limit, quantum confinement of the electronic wave functions gives rise to intriguing quantum mechanical effects. Transition metal dichalcogenides (TMDs) – a specific class of 2D semiconductors – exhibit a particularly unique example of this. Their optical properties are dominated by excitons, electron–hole pairs in semiconductors bound by the Coulomb force[40,41]. Reduced dielectric screening of the quantum-confined excitons leads to binding energies of hundreds of millielectronvolt, many times larger than the thermal energy at room temperature[42]. As a result, excitons are not dissociated into free carriers unlike for bulk semiconductors that are used for Mie resonant nanostructures (e.g. silicon). The resonant excitation of excitons gives rise to a very sharp and strong peak in the absorption spectrum, that can effectively be tuned over several 100 meV with changes in the index of the surrounding environment[43,44], electric/magnetic fields[43,45], strain[46,47], or electrostatic carrier injection[48–50]. Besides providing strong and highly tunable optical properties, there has been incredible progress in the synthesis, handling, and understanding of these 2D quantum materials[41,51]. This enables facile integration of these atomically-thin layers into more complex nanophotonic structures.

Based on these findings and developments, one can raise several interesting questions: *How thin can we make metasurfaces? Can excitonic resonances in 2D quantum materials be harnessed as a new type of atomic-scale resonance for light modulation and manipulation? Can we leverage their tunability to realize dynamically controllable and atomically-thin metasurface components?* The answers to these questions bring about a paradigm shift in the design principles and implementations of metasurfaces. The exciton resonance of monolayer TMDs is an intrinsic material property, i.e. a resonant response in the material's susceptibility to incident electric fields. As such, resonant light-matter interactions are warranted irrespective of the shape and location of the material in the metasurface geometry. This alleviates the conventional geometrical constraints that commonly confine the design of optically-resonant metasurfaces, and brings about new degrees of freedom in the metasurface design.

The aim of this tutorial article is to answer the above questions and to provide a general and didactical introduction to the field of atomically-thin optics. We outline the role of exciton resonances and the associated opportunities to develop a new class of tunable optical elements. To this end, the tutorial is structured as follows. In section 2 we briefly introduce the main concept of optical metasurfaces including the underlying resonant mechanisms. We also briefly review the state-of-the-art mechanisms for dynamic tuning of resonant metasurfaces. In section 3, we describe the relevant physics of excitonic materials, including light-matter interactions and mechanisms for exciton tuning. We include a detailed overview of available excitonic materials in which we compare their exciton related properties. In section 4, we discuss the three main applications of 2D materials in metasurfaces: (1) passive metasurfaces (no role for exciton), (2) hybrid metasurfaces in which excitons are coupled to optical resonances, and (3) tunable flat optical elements. Finally, in section 5 we discuss the outlook of excitonic materials for flat optics in terms of developments, limitations, and future opportunities. Throughout this tutorial article, we will refer to existing review papers that cover important aspects of the content in more detail.

## 2. Background: metasurface physics

### 2.1 Engineering amplitude and phase of resonant light scattering

The key ingredient of resonant metasurfaces is the ability to accurately engineer the amplitude and phase of the light scattered by a nanostructure. But what is resonant light scattering, and how can it be used to gain control over these parameters? To explore this, we will first briefly assess the basic principles of light scattering.

Light incident on a (nanoscale) piece of material induces local polarization of the material. Depending on the electronic band structure of the material, free electrons (in metals) or bound electrons (in dielectrics) are slightly displaced with respect to their equilibrium position. The resulting charge displacement induces a local dipole moment $\boldsymbol{p} = \varepsilon_0 \chi \boldsymbol{E}_i$. Here, $\varepsilon_0$ is the free-space permittivity, $\boldsymbol{E}_i$ is the local driving electric field, and $\chi = \varepsilon_r - 1$ is the material's electrical susceptibility, which quantifies how strongly the material responds to external electric fields. The total dipole moment of the particle can be obtained by integrating over the particle volume. Note that $\boldsymbol{E}_i$ corresponds to the *total* local electric field, which includes both the incident field as well as the fields induced by the scattered field of neighboring dipole moments. For a very small particle (diameter $d \ll \lambda/2\pi$) and small $\chi$, we can neglect the local contributions to $\boldsymbol{E}_i$ and assume a uniform electric field distribution (this is the *quasi-static approximation*). In this simplified scenario, the total dipole moment of the particle is simply $\boldsymbol{P} = \int \varepsilon_0 \chi \boldsymbol{E}_i dV = \varepsilon_0 V (\varepsilon_r - 1) \boldsymbol{E}_i$. For larger particles and strong susceptibility, the non-uniform local field distribution needs to be considered. The particle's response can no longer be described by simple dipole moment - retardation effects and high-order multipoles can play a significant role[52].

The electric field component of light can thus induce an electric dipole moment in the nanoparticle. The next step is to realize that the incident electric field is not static but rather oscillates at optical frequencies, described by a harmonic time-dependence $\boldsymbol{E}_i = \boldsymbol{E}_0 e^{i\omega t}$. As such, the total induced dipole moment in the nanoparticle will also oscillate at frequency $\omega$. This oscillating dipole will generate electromagnetic radiation that constitutes the scattered fields. Light scattering is thus the result of two steps: 1) incident light polarizes the material and induces a local dipole moment that oscillates at optical frequencies. This process takes energy from the light field and temporarily stores it in the oscillating charges. 2) The oscillating charges re-radiates the energy as scattered fields. Note that depending on the material's optical properties (described by the complex dielectric constant $\varepsilon = \varepsilon' + i\varepsilon''$), not all energy may be re-radiated. A fraction of the energy can be lost to heat in the nanoparticle due to absorption.

The oscillating electrons (either bound or free) experience a restoring force due to Coulombic attraction by the positively charged nuclei. This restoring force can give rise to a resonance condition where the charge displacement is maximized, analogous to a harmonic oscillator at resonance. The exact nature of the light-matter interaction that governs the resonant condition depends on the type of material (section 2.2). At resonance, the light-matter interaction is very strong and the optical cross sections (i.e. extinction and scattering cross sections) can be >10 times larger than the geometrical cross sections. As such, electromagnetic energy is "funneled" into the resonant nanoparticle and scattered with a large efficiency. The phase of the scattered light can take values between 0 radians - in-phase with the incident light (low-frequency side of the resonance), to $\pi$ radians (180 degrees) out of phase with the incident light (high-frequency side of the resonance). At resonance, the phase-lag is exactly $\pi/2$ radians (Figure 1a).

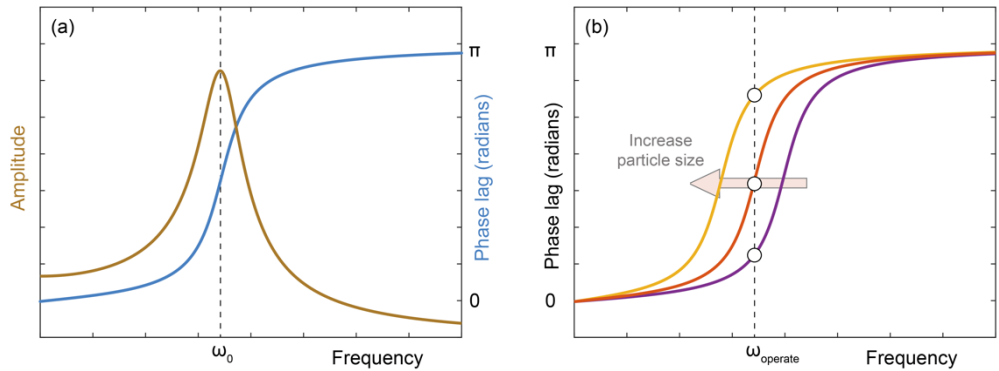

*Figure 1: Engineering amplitude and phase of resonant light scattering. (a) Amplitude (brown) and phase lag (blue) of the scattered light with respect to the incident light around the resonance frequency $\omega_0$. The amplitude peaks at resonance, while the phase-lag transitions from 0 to $\pi$ radians through the resonance. (b) Engineering the phase of the scattered light at the operating frequency by tuning the particle size (amplitude curves not shown).*

The resonance condition strongly depends on the geometry and dielectric environment of the nanoparticle. For example, larger particles exhibit a lower resonance frequency than smaller particles due to retardation effects. This scaling can be leveraged to achieve control over the amplitude and phase of the scattered light. If the incident frequency is fixed at the desired operating frequency, the geometry (or dielectric surrounding) of the nanoparticle can be varied to select the resonance amplitude and phase at will (Figure 1b). Such amplitude and phase engineering of resonant light scattering lies at the heart of optically-resonant metasurfaces. Note that in the basic discussion so far, the scattering amplitude and phase are "locked" and cannot be engineered independently. Also, the maximally achievable phase shift is in the [0- $\pi$] range whereas a full $2\pi$ phase range is desirable for high-efficiency metasurfaces. Advanced concepts like degenerate resonances[6,7], geometric phase, and propagation phase (discussed in section 2.3) provide more universal degrees of freedom at the expense of complexity or thicker layers.

## 2.2 Plasmon and Mie resonances

Fundamentally, two main resonant mechanisms can be distinguished as plasmonic (or free electron gas polarization) resonances and Mie (or dielectric displacement/polarization) resonances. For plasmon resonances, light couples to the resonant oscillation of the (geometrically confined) free electrons in metal nanoparticles. For geometrical Mie-type resonances, light is trapped and internally recycled in dielectric or semiconductor material. While both mechanisms are described by Mie theory for spheres in homogeneous media, the light-matter interaction underlying the dielectric constant of the material is fundamentally different (Figure 2). Plasmon resonances exhibit physical currents comprised of free electrons moving through the metal nanoparticle. Mie resonances on the other hand, are characterized by displacement currents of the bound electrons in the filled electron shells[53]. As a result, Mie resonances can be lossless (i.e. no optical absorption), in principle, while plasmon resonances always exhibit resistive losses due to electron scattering (Joule heating).

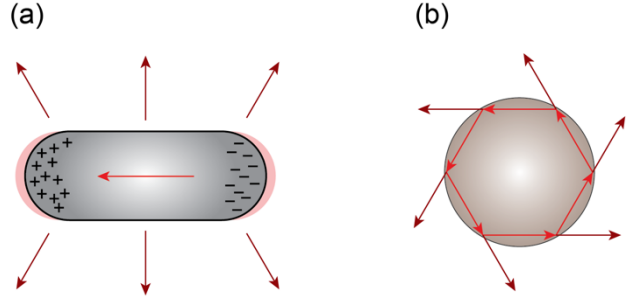

*Figure 2: Resonant light-matter interactions in nanoparticles. (a) Plasmon resonance in metal nanoparticle (left), where energy is stored in the resonant oscillation of the metal's free electrons. (b) Mie resonance where light is internally trapped and energy is stored in displacement currents.*

The different nature of the optical resonance has several important implications. First, due to electronic screening of electric fields inside the metal, plasmonic nanoparticles efficiently focus the near-field intensity on their surface. That is, the optical energy is confined to nanoscale volumes at the surface of the nanoparticles. This makes plasmonic nanoparticles very sensitive to the dielectric surrounding, enabling high sensitivity of change in the dielectric environment down to the single molecule[54] and sub-nanometer thickness levels[55,56]. For dielectric or semiconductor nanoparticles on the other hand, the optical near fields are mostly confined inside the particles volume[4,57]. Second, unlike plasmonic nanoparticles, dielectric (Mie) particles support magnetic resonances in addition to the electrical multipoles[53,57]. This provides a richer library of resonant modes to employ in nanophotonic systems. For example, by engineering the interference of light scattered by the electric and magnetic dipole modes the backscattering amplitude can be completely suppressed[58,59] – the so-called Kerker condition[60]. Finally, Mie-type resonances in semiconductor nanowires can be combined with photodetection schemes to realize resonant photocurrent generation in single nanostructures[61–64].

## 2.3 Propagation phase and geometrical phase

While resonant light scattering (described in section 2.1) enables accurate control over the scattering amplitude and phase, there are two important limitations. 1) The phase-lag induced by a single resonance is limited to the [0-$\pi$] range. 2) Every phase is intrinsically linked to a specific amplitude; these cannot be engineered independently. In a metasurface, elements with different phase-lags need to be combined to perform an optical function, which results in non-uniform amplitudes of the scattered fields. This deteriorates the metasurface efficiency. It is therefore desirable to be able to tune the phase over the full [0-$2\pi$] range with equal amplitude.

*Propagation phase*

Rather than employing resonant light scattering to control the local phase, metasurfaces based on propagation phase employ nanoscale waveguides that are oriented vertically on the surface (Figure 3a). The concept of propagation phase is derived from the phase pick-up imposed on a light field with free-space wavelength $\lambda_0$ propagating through a thin planar film with thickness $d$ and refractive index $n$: $\varphi = 2\pi nd/\lambda_0$. For a planar film with translational invariance, the optical pathlength $nd/\lambda_0$ is spatially uniform. However, by patterning the thin film into nanoscale vertical truncated waveguides, the refractive index can be controlled locally through the dimensions of the waveguide.

The overlap of the waveguide mode profile with the nanopillar material dictates the effective waveguide mode index $n_{wg}$: wider waveguides result in a higher mode index. The phase pick-up of light propagating through the nanoscale waveguides can thus be controlled locally. This enables polarization independent metasurfaces with very high and uniform transmission/scattering amplitudes. Additionally, the phase can be controlled in the full [0-$2\pi$] range by making the waveguides sufficiently tall. This highly flexible design has resulted in highly efficient metasurface lenses[65], but at the expense of significantly thicker nanostructured coatings in the range of 200-1000 nm.

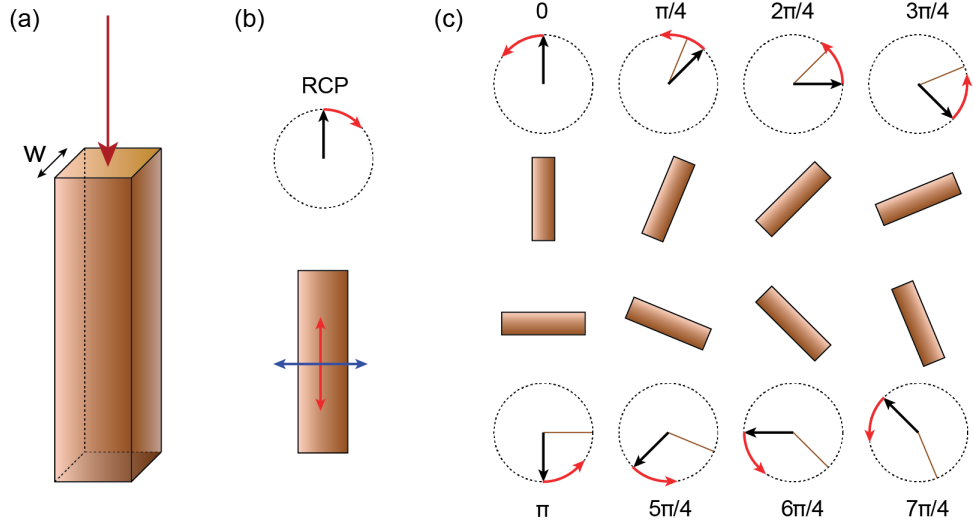

*Figure 3: Propagation and geometrical phase. (a) A nanoscale vertical truncated waveguide. Light propagates through the waveguide where the phase pick-up is governed by the waveguide length and mode index, which is engineered through the waveguide width (w). (b) Top view of a nanoscale half wave plate used in geometric phase: the scattered fields parallel to the nanoparticle's long axis (red) are engineered to have a π radians phase delay with respect to those scattered perpendicular to the long axis (blue). (c) Geometric phase delay (numbers in units of radians) and left-handed output polarization (black/red arrows) as a function of particle orientation (brown line). The incident light has right-handed circular polarization (RCP).*

*Geometrical phase*

Geometrical phase (also called Pancharatnam-Berry phase) employs standard resonant scattering elements and controls the phase of the scattered light through the orientation of the particle rather than the size (Fig. 3c). This enables all particles to be identical in size and therefore exhibit identical scattering amplitude. There are two prerequisites to the use of geometrical phase. First, the incident light must be circularly polarized. The phase-lag of the rotating electric field component of the incident light is imprinted onto the phase of scattered light. Metasurfaces are therefore selective for the handedness of the circular polarization. Second, the nanostructure shape and geometry need to be engineered such that it functions as a nanoscale half waveplate. That is, the scattered fields polarized parallel to the long axis of the nanoparticle (red arrow in Fig. 3b) are designed to have a $\pi$ radian phase delay with respect the scattered fields perpendicular to the long axis (blue arrow).

By rotating the coordinate system of the nanoparticle, the phase of the scattered light can be controlled. For example, when illuminated with right-handed circularly-polarized light (Fig. 3b), a waveplate oriented with its fast axis at an angle $\theta$ with respect to the incident electric field orientation will convert the polarization of the scattered light into the left-handed with a (geometric) phase delay of $2\theta$ (Fig. 3c)[66,67]. As such, a 0–180-degree rotation of the nanoparticle orientation results into a full [0-$2\pi$] range phase delay with equal amplitude. First introduced in 2001[66,68], geometrical phase is now widely employed in high-efficiency metasurfaces. It is oftentimes combined with propagation phase to achieve the required half waveplate response[69], enabling e.g. high-performance metalenses with diffraction limited focusing[70] and achromatic lenses with short focal length[71].

## 2.4 From resonant particle to metasurface

Sections 2.1-2.3 outlined the (resonant) engineering of the scattering amplitude and phase of single nanostructures. Metasurface employ judiciously designed sub-wavelength arrays of such nanostructures to shape the new wavefront, which results from interference of all scattered fields by the individual particles. A desired spatial phase profile (e.g. that of a lens) is discretized and each pixel is populated with a nanostructure that provides the desired scattering phase. The inter-particle spacing provides an additional degree of freedom in the metasurface design and can be leveraged to control the metasurface optical function. There are several design considerations to determine the optimal nanostructure spacing.

First, to ensure efficient interaction with all incident light, the particles spacing must be equal or smaller than the extinction cross section of the individual nanostructure. On the other hand, the particle spacing must be large enough to prevent (significant) near-field coupling. Typically, individual metasurface elements are initially designed without taking inter-particle interactions into account. While weak inter-particle coupling can be accounted for in the design procedure, strong interactions pose a challenge. When placed in very close proximity, strong coupling between two particles will cause the resonant modes to hybridize[72–74]. This induces strong shifts in the resonance condition, which deteriorates the metasurface performance.

Second periodicities in the nanoparticle pattern that are on the scale of the operation wavelength give rise to diffraction channels. Diffraction occurs for $\lambda_0/n < p$, where $p$ is the period of the nanoparticle array and $n$ is the refractive index of the medium. Depending on the metasurface functionality, diffraction effects may either be beneficial or detrimental. For example, metasurface anti-reflection coatings use deep-subwavelength periods to prevent diffraction effects in reflection[75,76]. Beam steering metasurfaces on the other hand, leverage diffraction effects to effectively redirect light in designed directions[16,18,77].

## 2.5 Dynamically-tunable metasurfaces

Thus far it is clear that the optical functionality of a metasurface hinges on the properties of its building blocks. Due to the static nature of the meta-units, once a metasurface is optimally engineered to match specific requirements, its structure, determined by shape, size and relative position of its constituents, is irreversibly fixed resulting into a rigid optical behavior. At the same time, post-fabrication tuning of the resonance phase and amplitude (Fig. 4a) is highly desirable for many novel applications including augmented reality, eye tracking, and optical communication. Over the past decades, diverse pathways have successfully led to actively-switchable and reconfigurable metasurfaces, each with its advantages but also limitations. For this tutorial, we briefly highlight the most common approaches.

*Phase change materials*

Phase change materials (PCMs) such as germanium-antimony-telluride compounds (GST), enable fast switching between the non-volatile amorphous and crystalline phases[78,79]. The rapid amorphous to crystalline switch of GST can be achieved via thermal annealing[80] as well as via optical or electrical pulses[81–86], whereas re-amorphization can be achieved via a melt-quench process[25]. PCMs are highly suitable for tunable metasurface applications in the mid-IR spectral range due to the high optical contrast between the two phases (large difference in refractive index)[87]. Incorporating PCMs in the resonant meta-unit (Fig. 4b) facilitates dynamic amplitude or phase tuning of the resonant properties for active metasurfaces[88–91]. Recent examples have demonstrate amplitude modulation of the scattered light intensity up to four-fold[26]. Moreover, the intermediate crystallization states can also be accessed, granting continuous tuning of the material properties such as post fabrication light-phase tuning within 81% of the full $2\pi$ phase range[92]. PCM metasurfaces offer large and continuous tunability and compact design schemes[93], but are limited to the IR spectral range, exhibit relatively slow switching speeds, and low power efficiency.

*Thermal tuning*

A second efficient approach to achieve dynamic modulation and spectral control of metasurfaces is via heating (Fig. 4c). Materials such as Ge and Si, along with their already prominent technological relevance, exhibit very strong thermo-optical coefficients: the refractive index changes with temperature. Controlling the temperature of a Si or Ge dielectric resonator would thus affect their optical resonances both in terms of spectral position and amplitude[94]. Thermal tuning via heating stages has already enabled reconfigurable Si metasurfaces[92,95], InSb metalenses[96] and various meta-optics[97]. An ulterior and recently arising thermal tuning option is light-induced heating[98]. Here, the self-induced optical heating of the absorber plays a crucial role in the thermo-optical effect enabling the transition from in-resonance to off-resonance illumination and vice-versa. Thermal resonance tuning is very simple and compatible with the visible spectral range, but is hindered by small tunability, slow switching, and power inefficient.

*Electrostatic doping*

Tuning via electrostatic doping provides active electrical control over the optical response of a metasurface. Typically, the meta-atom's geometrical resonance (plasmon or Mie) is optically coupled to a thin film with a carrier density-dependent refractive index in a composite 3D architecture (Fig. 4d). A metal back reflector functions as the gate electrode, onto which a gate oxide and tunable metal-oxide thin film are placed to complete the capacitor structure. The array of nanoparticles is positioned on top of the metal-oxide film, most commonly indium-tin-oxide (ITO), whose optical properties can be substantially altered by tuning the carrier density. This geometry is particularly suitable for plasmon resonances, as normally incident light can very effectively couple to gap plasmons supported by the metal-insulator-metal cavity[99]. The field profile of the gap plasmon is strongly confined inside the ITO layer, enhancing the sensitivity to the refractive index in the ITO. By applying bias voltages of opposite sign, the ITO can be switched from the depletion to the accumulation regime of carriers. This strongly affects the resonant response of the meta-atom and, by extension, of the metasurface. These changes are maximized by operating near the oxide plasma resonance in the IR spectral range. In addition, the carrier density can be tuned in a continuous fashion, enabling active electrical control over the amplitude, polarization and phase of the light scattered by the metasurface[32]. Alternatively, graphene-based metasurfaces leverage the dependence of graphene's optical properties on its carrier density to directly tune the plasmon response of graphene in the IR spectral range[100–103]. Overall, electrostatic doping of metal oxides and graphene allows for large and continuous tunability, fast switching speeds, and low power consumption, but its applicability is mainly limited to the IR spectral range.

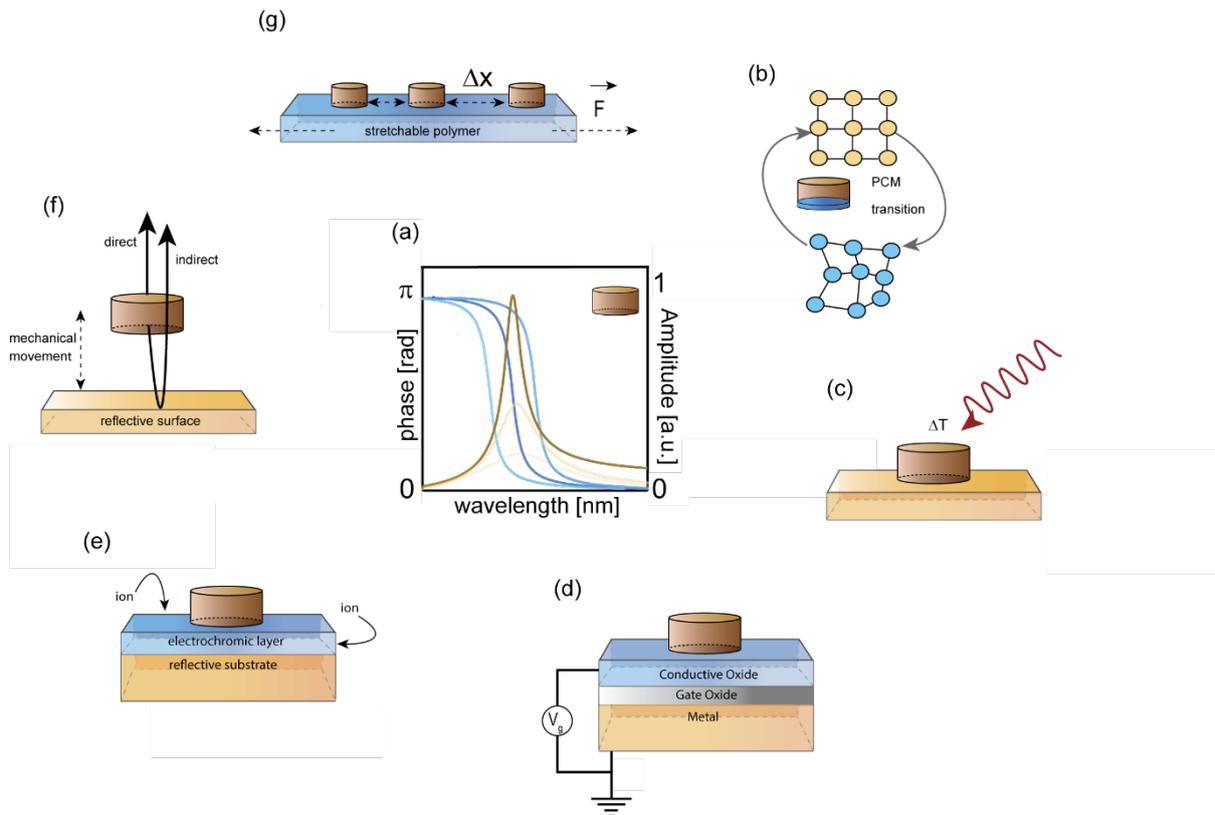

*Figure 4: Dynamic and reconfigurable metasurface mechanisms. (a) Active tuning of the amplitude (brown to yellow) or phase (blue) of the scattered fields of individual resonant elements is required for dynamic metasurfaces (examples b-e). Alternatively, active control over the (inter)particle position in an engineered dielectric environment can be used (examples f, g). (b) PCMs (blue) incorporated in a resonant particle change the resonance condition when switched between its crystalline (top) or amorphous (bottom) state due to a difference in refractive index. (c) Heating nanoparticles comprised of materials with a large thermal-optical coefficient shifts their resonance condition with temperature. (d) Electrostatic doping of thin metal-oxide (e.g. ITO) through gating schemes changes its refractive index in the IR spectral range. By optically coupling resonant nanoparticles to the ITO layer, the resonant condition of the nanoparticle can be tuned electrically. (e) Large changes in the refractive index of electrochromic materials can be achieved by ion intercalation. By overlapping the near-field of a resonant nanoparticle with the electrochromic layer, the resonance condition can be tuned continuously. (f) The resonance condition of nanoparticles placed above a reflective surface can be actively tuned by changing the distance above the reflector. Scattered fields that are reflected interfere with the directly scattered fields, which strongly depends on the position of the nanoparticle. (g) Changing the inter-particle distance in a metasurface using e.g. stretchable substrates provides direct control over the far-field interference of the scattered wave fronts, and thereby the optical function of the metasurface.*

(Electro)chemical intercalation

An alternative effective way to actively tune the resonance condition of (gap) plasmon resonators is via electrochemical intercalation[104] (Fig. 4e). The tuning here is granted by the incorporation of an electrochromic oxide such as $WO_3$ in the gap of the gap plasmon resonator[105], or by direct chemical hydrogenation of plasmonic particles[106,107]. For electrochemical intercalation, upon application of a bias with respect to a reference electrode, ions (e.g. $Li^+$, $H^+$) are reversibly injected into the electrochromic host material, thereby inducing large changes in the refractive index. Unlike for solid-state gating (Fig. 4d), these electrochemical reactions inject both an electron and an ion into the oxide. The electron is accepted by the host material, whereas the ion acts as a dopant that counterbalances the charge of the electron. As a result, electrochemical intercalation enables very large changes in the carrier density endowing active tuning of materials in the important visible spectral range. The slow yet continuous changes in the oxide refractive index enable the continuous and reversible shift of the plasmon resonance wavelength up to tens of nanometers, with tuned states holding stable for tens of minutes in open circuit conditions. Although tunable in the visible spectral range, electrochemical intercalation is very slow (moving ions), requires complex electrochemical cells, and exhibits finite stability due to degradation and unwanted chemical reactions.

Mechanical movement

Thus far, we have discussed how to achieve tunability by harnessing tunable, intrinsic material properties. Alternatively, microscale movement of miniature mirrors has historically been pursued to realize reconfigurable optical functions. Similarly, the optical response of a resonant nanostructure can also be impacted by its position in a tailored environment. In fact, optical resonances of high index semiconductors can be tuned both prior to fabrication, by manipulating their size and geometry[108], but also post-fabrication by bringing them in close proximity of a reflective surface (Fig. 4f).

To provide an example, the resonant response of a Si nanowire (NW) placed on quartz can be strongly tuned by varying distance of the NW above the reflective surface[28]. The standing wave formed by the incident and reflected light can modify the excitation efficiency of the different NW modes. Additionally, the presence of the mirror also impacts the scattered light, resulting from the interference between the light directly scattered by the NW and the light firstly reflected by the mirror surface. Combined, these effects impact both the light scattering intensity as well as resonance wavelength of the NW. Based on these insights, dynamic metasurfaces have recently been realized that employ micro-electrical mechanical systems (MEMS) technology that perform temporal color mixing and dynamic beam shaping[109]. Mechanical movement offers very large tunability in the visible spectral range, electrical control, and high switching speeds. However, the tunable structures require complex 3D architectures that are fragile and require additional packaging.

Stretchable materials

Rather than moving resonant elements collectively with respect to a reflective surface, metasurface functions can also be controlled actively by tuning the relative position of resonant nanoparticles within the metasurface. By changing the spacing between the metasurface elements, the far-field interference can be manipulated. The most intuitive example is that of high-index nanoparticles embedded into a low-index elastic membrane, i.e. a polymer such as polydimethylsiloxane[110] (Fig. 4g). Stretching the substrate provides direct control over phase profile of the scattered fields, hence altering its optical functionality. Following this rationale, multipurpose lenses with largely tunable focal length have been demonstrated[111,112]. Stretchable substrates offer very large and continuous tunability in the visible spectral range but require complex fabrication schemes and rely on slow mechanical activation.

**2.6 Summary**

In section 2 we have introduced the basic concepts of resonant light scattering and metasurface physics. Optical resonances form the foundation of metasurfaces by providing strong light-matter interactions and enabling control over the amplitude and phase of the scattered light at the nanoscale. Sub-wavelength arrays of resonant nanoparticles collectively perform an optical function through interference of all scattered fields. While highly efficient, metasurfaces based on plasmon and Mie resonances are intrinsically static with the optical function fixed after fabrication. The most common approaches to overcome this limit and realize dynamic tuning of optical resonances have been discussed. Successful tuning has been demonstrated, but these tuning mechanisms provide limited tuning, require complex 3D architectures, or consume significant power.

# 3. Background: Exciton Physics

Monolayer 2D materials offer the opportunity to further shrink a metasurface's thickness from hundreds of nanometers to atomically-thin (<1 nm). The key obstacle for such ultra-flat optic devices is achieving strong light-matter interaction in the atomically-thin media. Commercial photonic devices and optical elements enhance light-matter interactions by having thicknesses equal to or greater than the wavelength of light. However, atomically thin materials typically have thicknesses <1 nm which is less than 1/600$^{th}$ of the wavelength of red light. Therefore, resonant phenomena that can occur at ultrathin thicknesses, such as plasmons, phonons, or excitons are required to enhance the light-matter interactions for flat optics. Excitons in particular offer uniquely strong light-matter interactions. In this section, we introduce the basic concepts of excitons (and exciton-polaritons) which will be used in flat optics in section 4 of this tutorial article. Various other reviews in the literature have extensively covered the concept of plasmons and phonon-polaritons for resonant flat optics[113–115].

## 3.1 Semiconductor Band Structures

One of the defining discoveries of 20$^{th}$ century physics was the quantization of the energy of electrons in a hydrogen atom by Niels Bohr. Although Bohr merely studied the simplest system, a lone hydrogen atom, the quantization of electron's energy levels is also observed in semiconducting materials. The energy levels of an electron in a semiconductor depends on how it interacts both with the many different atomic nuclei and other electrons. Each allowed state for the electrons is defined by its energy and momentum, and each state can be occupied by two electrons due to spin degeneracy. When all states are plotted, they collectively form band diagrams (Figure 5a).

An electron can transition between any two states so long as it doesn't break the conservation of energy, momentum, and angular momentum (spin). If the two energy states in a transition have the same electron momentum, the electron can make the transition by absorbing or emitting a photon with the resonant frequency. The transition is then called a direct transition and is desired for its ability to produce highly efficient optoelectronic devices. However, if the two energy states do not have the same momentum, a phonon is also required to conserve momentum, and it is called an indirect transition.

In the band diagram there are two main classifications, the valence band where the electrons are bound to a single atom, and the conduction band where the electrons are not. In metals, these two bands overlap allowing electrons to flow easily, while in semiconductors an energy gap opens between the two bands. This requires electrons to be excited from the valence band before they can be in the conduction band. The size of this gap is called the band gap and is typically in the range of 0.5 to 4 eV. The band gap describes the lowest energy transition between the valence and conduction band. It can be further characterized into a direct or indirect band gap depending on its lowest energy transition. Materials with larger band gaps are referred to as insulators, but the distinction between semiconductors and insulators is arbitrary.

Band structures can be qualitatively understood using the Kronig-Penney model where the electrons are in a periodic array of square energy potentials that represent atomic nuclei[116]. However, quantitative studies of the band structures of materials are performed using tight binding models[117,118] or the more accurate, but computationally expensive, density functional theories[119,120]. Experimental measurements of a band structure are a very difficult endeavor, but can be performed using angle-resolved photoemission spectroscopy[121]. In the two-dimensional limit (i.e. for monolayer 2D semiconductors), the band structure is altered by quantum confinement effects that blue shift the band gap and typically leads to direct band gaps. For more details on band structures we refer to dedicated reviews and textbooks[122].

## 3.2 Excitons

When an electron is excited from the valence band to the conduction band, the now empty state in the valence band will be filled by another electron eventually. However, this process will create another empty state in the valence band, and the process will repeat. Instead of modelling these interactions by tracking every electron in the valence band, it is much simpler to only track the empty state as a quasiparticle called a *hole*. Since the hole is the absence of an electron, it behaves as if it has an equal and opposite charge of an electron as well as opposite spin as the excited electron. By definition, all of the interactions between the hole and other electrons in the valence band have already been considered, but the hole will also be bound to the excited electron through Coulomb interactions. The electron-hole pair is Coulomb bound to one another by the binding energy ($E_b$). Combined, the pair forms a new quasiparticle called an exciton that has zero net charge and spin. The exciton binding energy is therefore the energy needed to disassociate the electron-hole pair from one another.

As the electron is bound to a hole in the valence band, it is not free to move throughout the lattice. Therefore, it is not yet in the conductance band. Instead, the exciton occupies one of multiple states below the conduction band minimum (CBM). These new states below follow a Rydberg series[49] as shown in Figure 5a. Since the hole is in the valence band, the exciton wavefunction is still spatially concentrated around the atom the hole is associated with (Figure 5b). In this sense, the electron is said to have transversed the optical band gap, but not the electronic band gap. The optical band gap energy ($E_O$) is the energy difference between the valence band maximum (VBM) and the lowest exciton state ($n=1$). The electronic band gap ($E_e$) on the other hand, is the energy difference between the VBM and CBM. As the discrepancy between the optical and electronic band gap is caused by the Coulomb energy binding the electron and hole, the energy difference is equal to the exciton binding energy ($E_e - E_O = E_b$).

The binding energy of the exciton depends on both the band structure of the material as well as environmental characteristics, including the dielectric landscape, carrier density, and external electric or magnetic fields. Each of these parameters can be leveraged which can all be controlled to alter the properties of the exciton (see section 3.5). In bulk crystals, the exciton wavefunction extends in all three directions and there is a large dielectric environment that effectively screens the Coulomb interaction between the electron and hole, resulting in low binding energies (5-10 meV). In most common bulk semiconductors (Si and III-V materials), exciton binding energies less than the thermal energy at room temperature (26 meV). Therefore, excitons quickly dissociate and are not observed in bulk crystals except at low temperature.

Bulk van der Waals materials form an exception to this rule, where weak out-of-plane bonding confines the exciton wavefunction to two-dimensions[123]. Additionally, as the thickness of a bulk crystal is decreased, quantum confinement of the exciton wavefunction will occur for crystal thickness comparable to the size of the exciton wavefunction in bulk (typically ≈10 nm). The dielectric landscape also changes as the relatively high refractive index ($n > 1$) crystal is replaced with vacuum ($n = 1$), thereby reducing the Coulomb screening by the surrounding material. Combined, these effects result in a very strong increase in the exciton binding energy up to several hundred meV (~700 meV for monolayer $WS_2$)[124]. As such, excitons are highly stable in monolayer TMDs, even at room temperature.

Excitons possess an electric dipole moment as they are comprised of two oppositely charged particles. Therefore, excitons can only be excited when the electric field is aligned with the excitonic dipole moment. The excitonic dipole is oriented in-plane in atomically thin materials due to quantum confinement so excitons can only be excited by in-plane polarized light. Atomically thin semiconductors have large anisotropic optical properties[125] as a result, which can be used in polarizers, sensors, and lenses. Materials such black phosphorus[126], $SnS$[127], and $ReSe_2$[128] also possess in-plane anisotropy that are advantageous for flat optics due to their crystal structures.

In the context of this tutorial, excitons have garnered the interest of the semiconductor optoelectronics community due to their large dielectric susceptibility, which enables subwavelength photonics devices as well as to be post-fabrication tunability of their optical properties. The common method to model the contribution to the refractive index from an exciton was developed by Hendrik Antoon Lorentz in 1915 and as such is called the Lorentz oscillator model given by:

$$\epsilon(E) = 1 + \frac{fE_x^2}{E_x^2 - E^2 - i\Gamma_x E}$$

Where $\epsilon$ is the complex permittivity related to the complex refractive index ($\tilde{n} = n + ik$) by $\epsilon = \tilde{n}^2$, $E$ ($E_x$) is the energy of the incident light (exciton resonance), $f$ is the oscillator strength, and $\Gamma_x$ is the linewidth of the exciton. The physical interpretation of the oscillator strength is the probability of a resonant photon being absorbed. Since a photon being absorbed requires an electronic transition, the oscillator strength can be calculated directly using Fermi's Golden Rule[129,130]. The oscillator strength is a unitless quantity, but it is common practice to report $fE_x^2$ as the "oscillator strength" resulting in it having units of energy squared. The linewidth encompasses all possible damping processes the exciton can undergo such as radiative decay or scattering with phonons, electrons, defects, or other excitons. The oscillator strength, linewidth, and exciton energy can all be measured experimentally using spectroscopic ellipsometry[131] or reflection and transmission measurements[132].

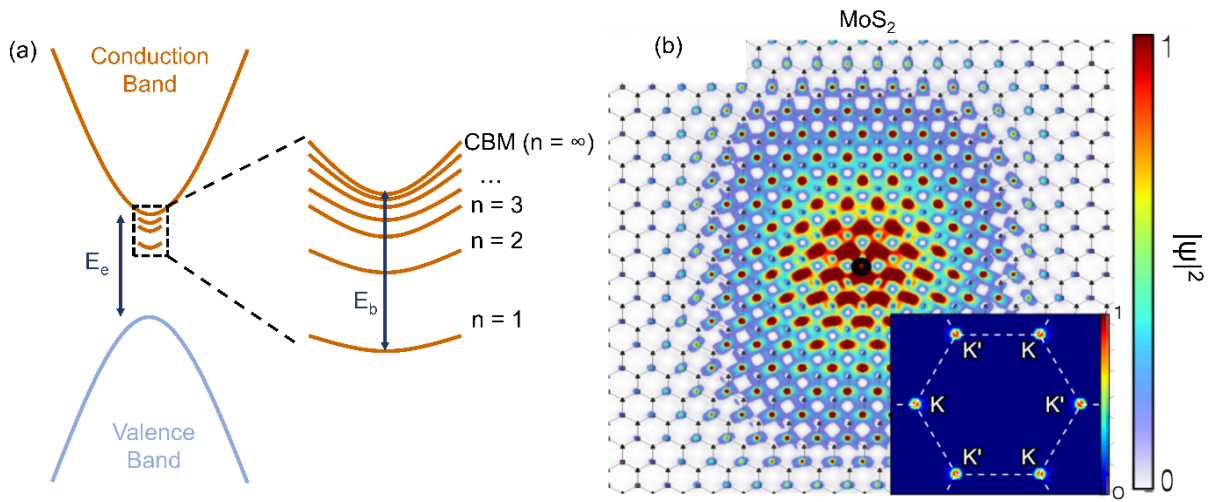

*Figure 5: Band structure of semiconductors and wavefunction of excitons. (a) Band structure of a direct bandgap semiconductor with the electronic band gap ($E_e$) labelled, and a diagram of the Rydberg series of excitons where the binding energy ($E_b$) is the energy difference between the ground state (n = 1) and the conduction band minimum (CBM). (b) The square of the wavefunction (ψ) of the primary exciton in monolayer $MoS_2$ in real space showing that the exciton is concentrated around the Mo atoms. The inset depicts the square of the wavefunction in k-space. Figure 5b is reproduced with permission from ref [133].*

## 3.3 Excitonic materials

Excitonic materials are characterized by three important parameters: their type of band gap, binding energy, and optical band gap. The type of band gap affects the efficiency of the semiconductor in optoelectronic applications including LEDs and photovoltaics, since direct band gaps exhibit higher quantum yields than indirect ones[134]. The binding energy is important since it determines the stability of the exciton. A large binding energy is also indicative of a high refractive index since both values depend on the joint density of states[135–137]. In addition, the binding energy determines the temperature range at which stable excitons form. Finally, the optical band gap determines the energy range in which excitons interact with light, and therefore, in which range the exciton can be used in applications.

All three characteristics are shown in Figure 6a for common atomically thin materials in their 3D (multi-layer) and 2D (single layer) form. The most commonly used bulk semiconductors are also shown. The data in Fig. 6a corroborates all trends that were discussed in sections 3.1 and 3.2. First, bulk semiconductors such as Si and III-V materials have binding energies <26 meV preventing the observation of excitons at room temperature. Second, all semiconductors exhibit a direct band gap in the monolayer limit except for InSe. Third, in the transition from 3D to 2D crystals, the band gap increases slightly, and the binding energy increases strongly to 100's of meV.

Figure 6b shows the relation between change in binding energy when going from bulk 3D to 2D crystals and the refractive index of the semiconductor at the optical band gap energy. A general positive correlation can be observed that indicates that the change in binding energy is larger for semiconductors with higher refractive indices. The dielectric landscape (and thereby screening) will be larger for high-index semiconductors. All of the semiconductors in Fig. 6b host Wannier-Mott excitons, which typically show a binding energy that scales as $E_b \propto \frac{1}{n^4}$[138]. However, the predicted $\Delta E_b = E_b^{2D} - E_b^{3D} \propto \left(1 - \frac{1}{n^4}\right)$ relation is not observed in Fig. 6b since the binding energy is also affected by changes in the band structure. Moreover, the refractive index of the semiconductor is not constant with thickness. The molybdenum TMDs also appear to show smaller changes in the binding energies than expected based on the other materials. An ongoing question regarding these materials is a large discrepancy between their theoretical and observed binding energies[139]. Understanding the relation between the binding energy, refractive index, and thickness required deeper investigation of calculations and estimates of excited state, many-body band structures of strongly confined solid-state systems which are difficult and computationally expensive to perform and therefore a frontier area of research[133].

Altogether, Fig. 6 functions as a frame of reference to identify the optimal excitonic material for nanophotonic structures and devices.

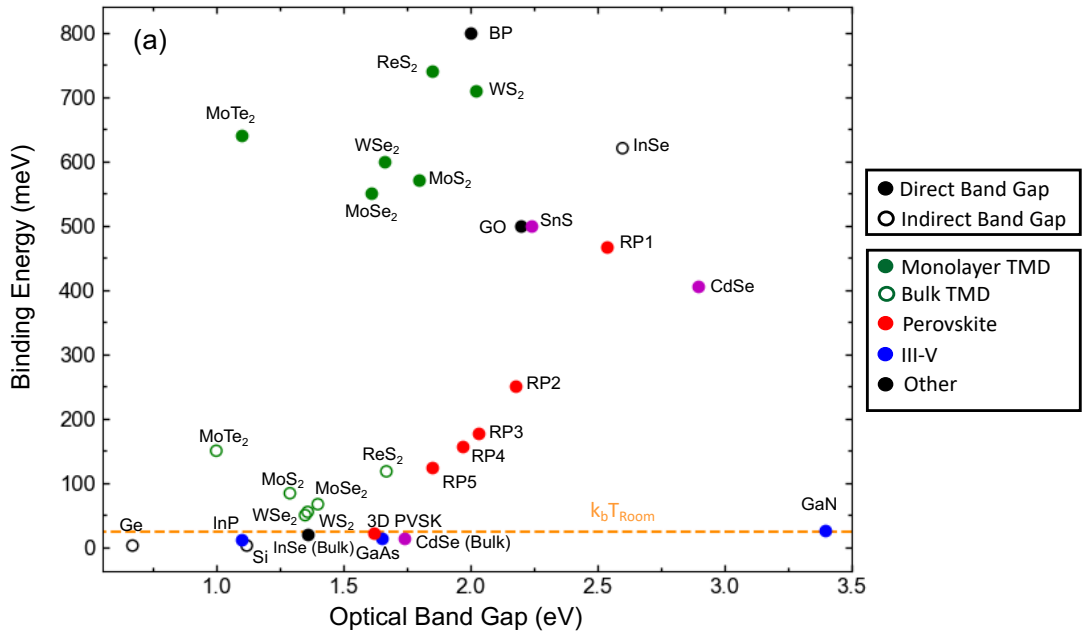

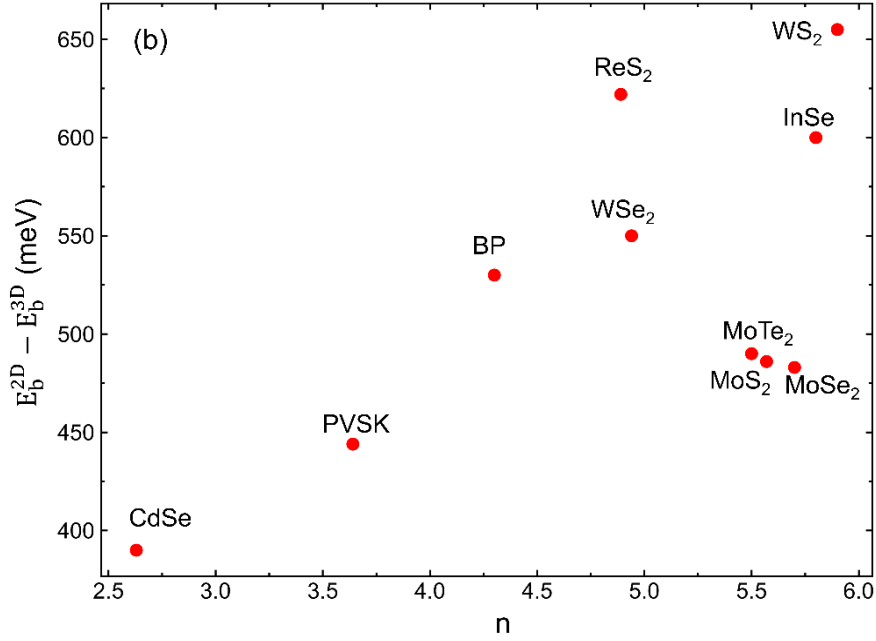

*Figure 5: Excitons in 3D and 2D semiconductors. (a) Binding energy and optical band gap of common atomically thin semiconductors in both their bulk and monolayer forms. Values for the most common bulk semiconductors are also shown. The legend indicates both what type of band gap the semiconductor exhibits and what family of materials it belongs to. All of the values are taken from literature[124,140–180]. TMD is transition-metal dichalcogenide, PVSK is perovskite. RPX is Ruddlesden-Popper phase perovskite of X layers thickness. (b) The difference in binding energy between 2D and 3D crystals as a function of their refractive indices for light resonant with the primary excitons. A general positive correlation can be observed between the two parameters since the dielectric screening will be reduced by more in the 2D limit for high refractive index semiconductors. The values for the refractive index are from literature[181–185].*

### 3.4 Tuning Excitons

Large refractive index values near the exciton resonances enable ultrathin photonics with extraordinary properties. However, active tuning of the material's optical properties is required for novel applications including sensors, modulators, and phased arrays. Two-dimensional excitons possess outstanding tunability for these applications[50,186,187] since their properties depend on the local Coulomb environment and crystalline structure. The tunability of 2D semiconductors has already been shown to be orders of magnitude larger than today's photonic materials such as silicon[188,189] and III-V materials[190]. Here, we will discuss the most common methods to tune the exciton properties.

<u>Magnetic fields</u>

The effects of magnetic fields on the electronic structure of matter have a long history as the 2$^{nd}$ Nobel Prize in Physics in 1902 was jointly awarded to Pieter Zeeman and Hendrik Lorentz, for their discovery and explanation of what would come to be known as the Zeeman effect. The Zeeman effect[191] is the observed splitting of electron energy states in the presence of external magnetic fields, and a theoretical description of the Zeeman effect required the development of quantum mechanics and electron spin. The electron spin acts as its own magnetic moment ($\boldsymbol{\mu}$) whose potential energy (U) in an external magnetic field ($\boldsymbol{B}$) depends on the alignment between the magnetic moment and field ($U = -\boldsymbol{\mu} \cdot \boldsymbol{B}$). Therefore, an energy state will split into two separate states where the lower (higher) energy state is for an electron spin parallel (antiparallel) to the magnetic field. The two new states' energies will be separated by $2\mu B$, and they will be centered around the original state's energy. For excitons, the Zeeman effect can be most easily observed in atomically thin crystals that lack inversion symmetry such as odd-layered TMDs since their electron band structures depend on spin[192,193]. However, a major drawback in using the Zeeman effect for modulation is that magnetic fields of ~1 T induce energy changes of only a few meV[43,45]. Therefore, the Zeeman effect requires magnetic fields that are difficult to produce in practical application settings, and the energy shift is only significant at cryogenic temperatures where the linewidths of excitons are extremely small and limited by intrinsic disorder only.

Electric fields

The electric analog of the Zeeman effect was discovered by Johannes Stark in 1919. In parallel to the experience of Zeeman, the effect was named the Stark effect and earned the 1919 Nobel Prize in Physics. In a similar fashion to the Zeeman effect, the Stark effect causes an energy state to split into two separate states due to the interaction between an external electric field and the exciton's electric dipole[194]. The simplest experimental setup produces an out-of-plane electric field that is perpendicular to the exciton dipole in monolayers minimizing the Stark effect. The obvious way to overcome this obstacle is to produce an in-plane electric field, but these fields are normally more difficult to produce than out-of-plane ones. Another possible approach is to use bilayer heterostructures that host interlayer excitons[195,196]. As the electron and hole will be located in two different atomic layers in this heterostructure, the exciton dipole moment will be shifted out-of-plane enhancing the Stark effect[195]. Stark effect photonic modulators are still uncommon in excitonic devices since they also require large electric fields to produce small energy shifts[197,198], similar to the Zeeman effect.

Electrostatic gating

One of the most common modulation techniques is electrostatic gating. In electrostatic gate design, a voltage is applied between a back gate electrode and the semiconductor in a capacitive configuration. The gate voltage changes the Fermi level in the electrically connected semiconductor layer and thereby the carrier density. If the Fermi level is shifted towards the conduction (valence) band, the number of free electrons (holes) will change. A change in free carrier densities affects the excitonic properties through Coulomb screening, Pauli blocking, and scattering[50]. Of these three mechanisms, Coulomb screening typically affects the exciton the most as increased Coulomb screening reduces the binding energy and oscillator strength of the exciton while also broadening its linewidth. Pauli blocking acts to blue shift the band gap as states at the conduction band minima are now occupied. Scattering increases the linewidth since it adds a pathway for the exciton to dissipate energy. Combined, electrostatic gating can be used to completely suppress the exciton transition.

Electrostatic gating is particularly effective for monolayer 2D materials, as the entire volume of the material is within the electrostatic accumulation/depletion layer. Gating of monolayer TMDs has demonstrated large changes in the exciton amplitude and energy[48,49]. One of the main hurdles for electrostatic gating is minimizing the contact resistance with the semiconductor as a high resistance prevents the injection of free carriers. However, contact resistance can be minimized through contact engineering for the specific material and is an area of active research for atomically-thin 2D semiconductors[199]. Electrostatic gating will result in complicated designs in metasurfaces that will require contacts for every semiconductor island[200].

Strain and temperature

Up to this point, the three exciton tuning techniques that have been discussed have all been electromagnetic in nature. However, the band structure, and as a result the exciton, can be altered by physical processes such as strain and temperature as well. Strain modulation has attracted significant interest recently as atomically thin optoelectronics could enable wearable devices in the future. When a material experiences strain, the lattice deforms altering the band structure of the material. Normally, this results in a red shift in the exciton energy for positive strain as the lattice spacing increases[47]. Strain can also change two other aspects of the excitons. First, different points in momentum space are affected more than others making it possible for the fundamental transition to change. This has been observed in TMDs that undergo a transition from a direct band gap to indirect for a strain of ≈2%[201]. Strain can also reduce the exciton binding energy by altering the free carrier density[201]. This occurs when the strain changes the curvature of the conduction band minimum. When these bands become flatter (narrower), more (less) electrons will be thermally excited to the conduction band allowing them to Coulomb screen the exciton. This reduces the binding energy and consequently lowers the exciton amplitude. The band structure can also be modulated by temperature since the thermal expansion coefficient results in a change in the lattice structure[202]. In this sense, thermal modulation is similar to strain-induced modulation. However, the main difference is that thermal modulation can also change the free carrier density since a change in thermal energy changes the amount of thermally excited free carriers.

## 3.5 Exciton-Polaritons

Excitons strongly enhance resonant light-matter interactions and are intrinsic electronic resonances of the atomic lattice itself. However, light-matter interactions can be further enhanced when the exciton is in the presence of a resonant cavity with a similar energy ($E_c$). The cavity mode can either be external, where the excitonic material is placed in a resonant light-confining structure that forms the cavity mode, or internal, where the excitonic material is a part of the medium forming the cavity mode (e.g. Mie mode). In the presence of a cavity mode, energy will be exchanged between the photons in the cavity and the excitons in the semiconductor. This interaction can be modelled using the Jaynes-Cummings Hamiltonian ($H_{JC}$):

$$H_{JC} = H_x + H_c + H_{int} = \begin{pmatrix} E_x & g \\ g & E_c \end{pmatrix}$$

Where $H_x$ ($H_c$) is the Hamiltonian of an isolated exciton (cavity photon), and $H_{int}$ accounts for simplest energy transfer between the exciton and cavity, i.e. the exciton absorbs a photon from, or emits a photon into the cavity. In the matrix form of the Jaynes-Cummings model, the wavefunction is a 2 × 1 matrix, $\Psi = \begin{pmatrix} \alpha \\ \beta \end{pmatrix}$, here $\alpha$ ($\beta$) represents the exciton (photon) component of the particle. The two coefficients are known as the Hopfield coefficients.

The meaning of the coupling parameter, $g$, can then be interpreted as the rate of energy transfer between the exciton and photon states. The coupling parameter appears as the off-diagonal terms in the Hamiltonian. If the coupling parameter is larger than linewidths of the exciton and cavity modes, energy will be transferred between the two states with minimal energy loss through other means. In this case, the wavefunction will oscillate between the exciton and cavity modes on typical time scales of tens of femtoseconds. On times scales much larger than tens of femtoseconds, the wavefunction will act as a quasi-particle that is part exciton and part photon, called an *exciton-polariton*.

Exciton-polaritons are advantageous over pure excitons since they exhibit higher mobilities, can achieve near-unity absorption, and enable absorption below the band gap. The exciton-polariton peaks are observed at the eigenenergies of the Jaynes-Cummings Hamiltonian, and since the Hamiltonian is a 2 × 2 matrix, there are two different exciton-polariton peaks. The two peaks are called the upper (UEP) and lower (LEP) exciton-polaritons, and they are labeled by their higher and lower energies, respectively. When the exciton and cavity modes are resonant with one another, the exciton-polaritons are 50% exciton and 50% photon, and the energy differences between the unperturbed exciton energy and the exciton-polaritons is equal to the coupling parameter. Although the coupling parameter characterizes the coupling, it is common practice to report the coupling strength of the system using the Rabi splitting value ($\hbar\Omega_{Rabi} = 2g$) that is equal to the energy difference between the UEP and LEP when the cavity mode is resonant with the exciton. However, both the coupling parameter and Rabi splitting characterize the light-matter coupling equally well.

## 3.6 Coherent exciton radiation

The strong light-matter interaction offered by exciton resonances translates into strong signals in photoluminescence (PL) measurements as well as large contrast in reflection spectroscopy. For both measurements, the amplitude and line shape of the exciton resonance as observed in the experimental spectra is dictated by the material's quantum efficiency and decay rates. However, the role of coherence in the exciton radiation is fundamentally different.

In photoluminescence experiments, an electron is typically excited high into the conduction band by absorbing a high-energy photon ($E > E_x$). This excitation event is immediately followed by thermalization of the electron to the bottom of the conduction band, and finally into a bound exciton state. In the thermalization process the electron loses all phase information from the excitation field, which corresponds to a loss of coherence. The exciton can then decay through the emission of a photon (radiative decay) or lose its energy non-radiatively to phonon modes (non-radiative decay). Both decay channels are characterized by a rate, $\gamma_{rad}$ and $\gamma_{nr}$ respectively. The relative contribution of the $\gamma_{rad}$ to the total decay rate $\gamma_{tot} = \gamma_{rad} + \gamma_{nr}$ defines the quantum efficiency: $QE = \gamma_{rad}/\gamma_{tot}$.

When the exciton is generated by absorption of a photon with an energy exactly equal to the exciton energy, thermalization does not occur, however. Such *resonant excitation* gives rise to the coherent excitation of excitons. If the exciton's intrinsic coherence time is sufficiently long, light can be re-emitted without losing coherence with the incident light. This results into fully coherent light scattering where the phase delay of the scattered light is governed by the susceptibility of the material. The exact phase relation with the incident (non-scattered) light dictates the exciton line shape through interference. Such coherent light-matter interaction offered by excitons provides the opportunity for metasurfaces to employ exciton resonances as a new building block in wave front shaping (see section 4).

Analogous to incoherent PL, non-radiative decay channels for the exciton such as phonon scattering also negatively impact the exciton population in coherent scattering and give rise to broadening of the spectral line shape. Recent works have shown that the interference of the exciton radiation with a cavity reflection offers the additional opportunity to extract a pure dephasing rate[203,204]. This quantum mechanical analysis of the 2D nature of the exciton radiation captures the finite coherence time and distinguishes between loss of coherence due to non-radiative decay and due to pure dephasing. While the exciton density is not affected under pure dephasing, the coherence of the exciton is lost. This also results into broadening of the excitonic line width.

## 3.7 Summary

In section 3 we have discussed the fundamental aspects of exciton physics. Excitons provide a uniquely strong light-matter interaction and are tunable over 100's of meV using external stimuli. The resonant response is intrinsically linked to the electronic band structure and thus does not require nanopatterning at the length scale of the optical wavelength. Due to quantum confinement and reduced screening, excitons have very large binding energies in monolayer TMDs and are stable at room temperature. Coupling excitons to optical cavities further enhances the strength of light-matter interactions through exciton-polaritons. Combined, these features make excitons a highly promising new building block to manipulate light in novel metasurfaces.

## 4. Excitons in 2D metasurfaces

In this section, we will discuss the main focus of this tutorial: excitons for atomically-thin flat metasurface optics, in three parts. First, we briefly discuss the implementation of 2D (TMD) materials in flat optical elements that leverage the high refractive index of TMDs: *passive 2D dielectric metasurfaces*. Second, we show how exciton resonances in TMDs can couple to geometrical resonances in plasmonic or dielectric metasurfaces: *hybrid metasurfaces*. Third, we discuss the first demonstrations of active tuning of exciton resonances in nanophotonic devices and metasurfaces: *tunable flat optics through exciton resonance tuning.* Note that nanophotonic applications of TMD monolayers are far from limited to the metasurface flat optics discussed here. We refer the reader to comprehensive review articles from literature precedent that discuss coupling of excitons to individual Mie and plasmon resonant nanoparticles[205].

### 4.1 Passive 2D dielectric metasurfaces

Dielectric metasurfaces employ Mie resonances, propagation phase, and/or geometric phase to tailor the phase and amplitude of the scattered light, as outlined in section 2. The efficiency of dielectric metasurfaces strongly depends on the contrast in refractive indices between the nanoparticle and surrounding materials. Large index contrast causes the interfaces to be highly reflective, and thereby light can be trapped more efficiently inside the nanoscale cavity modes (Mie resonances)[206]. As such, silicon is one of the most popular materials for metasurface design due to its high refractive index of $n \sim 3.5$ in the visible spectral range[207]. Building on these insights, TMDs are readily recognized as attractive materials for dielectric metasurfaces. For wavelengths beyond the optical bandgap ($E_{photon} < E_o$), TMDs exhibit remarkably high refractive indices $n = 4 - 6$ without optical absorption[131,132]. The resulting strong light-matter interaction has been leveraged to demonstrate various passive dielectric metasurfaces by patterning multilayer TMDs[208,209], and even small-area monolayer flakes[210]. While promising, these implementations of 2D semiconductors in flat optics ignore the role of excitons in the optical functionality, and therefore are outside the scope of this tutorial's main discussion.

### 4.2 Hybrid metasurfaces

Although localized resonances in passive metasurfaces enable accurate control over the metasurface's optical properties, coupling of resonant cavity modes to other resonances including excitons provides additional control and degrees of freedom. By combining the strong light-matter interaction of metallic or dielectric metasurfaces with that of the exciton resonances, uniquely strong and controllable light-matter interactions occur. When the coupling between two resonant modes is weak, the electric field energy is exchanged relatively slowly between the exciton and cavity modes and dissipates out of the system through other means (through absorption or scattering). The maximum phase modulation for weakly coupled system is <180º as a result, similar to a single resonance (section 2.1). For strong coupling on the other hand, the exchange rate of the stored energy is larger than the loss rate of the individual resonators. The rapid circulation of energy between the two resonant modes forms new hybridized modes and allows for a large accumulation of phase (up to 180º) and potentially large absorption. The coupling strength can be controlled locally in a hybrid metasurface by engineering the cavity mode frequency and line width.

<u>Coupling to plasmonic metasurfaces</u>

One approach to strong light-matter interactions in a hybrid metasurface is by patterning a metal to create plasmon-polaritons and then placing a 2D semiconductor in its proximity. The exciton couples to the plasmon-polariton to form an exciton-plasmon-polariton (EPP). Hu et al. demonstrate such an approach by patterning rectangular holes into an Au thin film and placing a monolayer WS$_2$ on top to form a hybrid metasurface (Fig. 7a)[211]. The orientation of the rectangles is rotated between adjacent holes causing variations in coupling throughout the metasurface. In this metasurface, the right-hand circularly polarized (RCP) and left-hand circularly polarized (LCP) light accumulate opposite phases at each point of the metasurface. As each point can be treated as a wavelet from Huygen's principle, RCP and LCP refract in opposite directions as shown in Fig. 7b.

The ability to achieve large phase differences between adjacent elements of a metasurface is of upmost importance to the optoelectronics community since it can be used to create phased arrays for beam steering and metalenses that focus at the diffraction limit. A comprehensive study of the coupling between excitons and plasmon-polariton using a square array of silver nanoparticles has been done in the past[212,213]. It was observed in that study that the light-matter interaction strength depends heavily on the geometry of the metallic array since the geometry affects the radiative loss rate (scattering), allowing for metasurfaces to be made in both the weak and strong coupling regime. Likewise, excitons have also been shown to couple to dipole and quadrupole resonances using gold nanorods[214] as well as anapoles using a split cylinder metallic structure[215] that can achieve large Rabi splitting values.

Coupling to dielectric metasurfaces

In a similar approach of coupling excitons to plasmon-polaritons, strong light-matter interactions can also be formed by placing a semiconductor on top or beneath a passive dielectric metasurface. The exciton will then couple to the cavity mode to form exciton-polaritons. This approach can be favorable since it avoids the optical losses of metals in photonic devices, and it can be easily integrated into existing silicon devices. For example, Chen et al. fabricate a cylindrical hole pattern in SiN with monolayer $WSe_2$ placed on top (Fig. 7c), and demonstrate strong light-matter coupling to form exciton-polaritons[216]. The resulting far-field photoluminescent radiation pattern of the monolayer $WSe_2$ is altered by the formation of exciton-polaritons since these affect the electric field distribution within the $WSe_2$. The metasurface is then shown to alter the far-field emission, and that the far-field radiation is highly dependent on the symmetry of the metasurface (Fig. 7d). When changing from a square array of cylindrical holes to a hexagonal one, it is found that the far field radiation pattern changed from one with four-fold symmetry to one with six-fold symmetry which is also shown in Figure 7d.

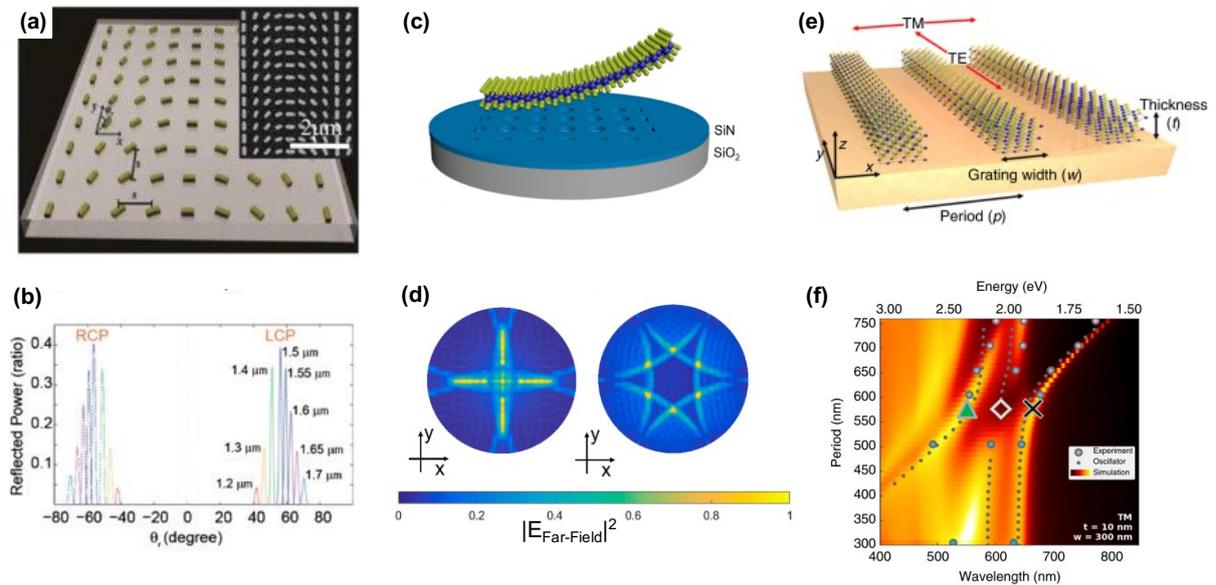

*Figure 7: Exciton-polariton and exciton-plasmon-polariton hybrid metasurfaces.* (a) A Pancharatnam-Berry Phase metasurface that forms a cavity mode that is coupled to a $WS_2$ monolayer to create an exciton-polariton. This metasurface results in right and left circularly polarized light being transmitted at different angles (b). Images were reproduced with permission from ref.[211] (c) Exciton-polariton hybrid metasurface along with its far-field radiation pattern (d). The far-field radiation maintains the symmetry of the metasurface patterning. Images were reproduced with permission from ref.[216] (e) A *metasurface fabricated by patterning a grating structure into thin film WS₂. Exciton-plasmon-polaritons are formed since the metasurface contains both excitonic and plasmonic materials. This metasurface was able to achieve extremely strong light-matter coupling for a 2D system as the energy separation between polariton branches was 410 meV (f). Images were reproduced with permission from ref.[217]*

Hybrid TMD metasurfaces

Up to this point, we have only discussed metasurfaces where the exciton was coupled to an adjacent photonic mode. However, exciton-polaritons can also be formed in metasurfaces by patterning the semiconductor directly. The photonic and matter resonances occur in the same material, further enhancing their coupling strength.

In a recent work[217], this was demonstrated by patterning a grating structure into WS₂ to form EPPs (Fig. 7e). Through the coupling of three resonances (exciton, plasmon, and cavity), near unity absorption was observed in WS₂ samples with thicknesses of ≈15 nm[217]. A characteristic of strong light-matter coupling is the splitting of an exciton absorption peak into two polariton branches. Most exciton-polaritons have Rabi splitting values in the range of 80 to 200 meV, but the WS₂ grating shows a peak splitting of 410 meV. The large energy separation in the WS₂ grating is a result of the exciton coupling to plasmon-polaritons causing the middle polariton branch to disappear at the anti-crossing point. The resonances of the WS₂ were controlled by the flake thickness, the grating period, and the grating width. As such, multiple parameters can be engineered to control the light-matter coupling.

Although this work focused on light polarized perpendicular to the gratings, hybrid metasurfaces can enhance the performance of other designs. In a second example[218], a waveguide in thin film WS₂ is created by etching circular holes in a square array. Strong light-matter interactions within the metasurface are observed without the need for a reflective substrate, which is typically used to enhance the light confinement. Although WS₂ exhibits little photoluminescent (PL) emission for more than one layer, this example also demonstrates that exciton-polariton in metasurfaces can be used to enhance the PL of an adjacent material by depositing monolayer WSe₂ on top of the metasurface and observe an 8x improvement in PL intensity. Other examples of hybrid metasurfaces include patterned suspended WS₂ films[218], Mie resonances in TMD nanodisks[219], and lens patterns to achieve direction exciton emission[220].

It is worth mentioning that subwavelength thick semiconductors can form exciton-polaritons and exciton-plasmon-polaritons without the need of patterning them into metasurfaces. By varying their thickness[221–223] or integrating them into metamaterials[224,225] with out-of-plane rather than in-plane periodicity hybridized modes can be achieved. Although these systems are not the focus of this tutorial, they can display similar phenomena as metasurfaces. However, their main drawback compared to metasurfaces is that they do not possess the potential for active wave front shaping since the optical response cannot be controlled locally.

## 4.3 Tunable flat optics through active exciton tuning

Besides (strong) coupling to resonant cavities, the resonant light-matter interaction of excitons in monolayer 2D materials itself can also function as a new building block for metasurface flat optics in the visible spectral range. The exciton's strong sensitivity to electrostatic doping is of particular interest as it has demonstrated to result into large changes in the exciton oscillator strength[50] and is compatible with well-established electronic platforms and fabrication technologies. Active tuning of exciton resonances in nanophotonic devices and dynamic metasurface flat optics has opened a new paradigm of metasurface design. The exciton response is intrinsic to the material and does not depend on the location and nanoscale geometry of the material. Here, we will discuss three pioneering examples of nanophotonic devices and metasurfaces for wave front shaping that employ active tuning of exciton resonance to achieve tunable optical functions.

Electrically-switchable atomically-thin mirror

Although the optical pathlength for light interacting with a monolayer TMD is very small, high quality materials that operate in the limit where the exciton decay is fully radiative (this is the *radiative limit*) can function as a perfect mirror, in principle[226]. Judicious engineering of the dielectric environment can also enhance the light-matter interaction. Combined with electrostatic doping, drastic changes in the reflectivity of a monolayer TMD can be achieved (Fig. 8a), thereby demonstrating an electrically-switchable atomically-thin mirror.

In two separate pioneering works, Scuri et al.[203] and Back et al.[227] demonstrate electrical tuning of efficient mirrors comprised of a single monolayer of $MoSe_2$. By placing a monolayer flake on an oxidized silicon wafer, the oxide functions as a resonant Fabry-Pérot cavity for light to enhance the light-matter interaction with the exciton resonance. At the same time, the silicon substrate functions as a back gate to control the electron density in the monolayer flake. At cryogenic temperatures (~4 K), the non-radiative decay channels for the exciton are suppressed and the exciton decay is fully radiative. For fully encapsulated flakes of $MoSe_2$, the resulting exciton oscillator strength is so large that the strong exciton radiation coherently interferes with the bare cavity reflection to produce asymmetric Fano line shapes[227] with peaks in the reflectivity up to 80%[203]. Electrostatic gating of the $MoSe_2$ flake completely quenches the exciton resonance and suppresses the reflectivity, facilitating electrical switching of the mirror function (Fig. 8b).

While only demonstrated on small non-patterned micron-scale flakes at cryogenic temperatures, these pioneering examples demonstrate how monolayer materials can achieve high optical efficiencies with electrical control over the optical function. More recently, follow-up experiments with electrically-tunable light scattering by a monolayer flake at cryogenic temperatures have demonstrated electro-mechanical modulation[228] and small-angle beam steering[229].

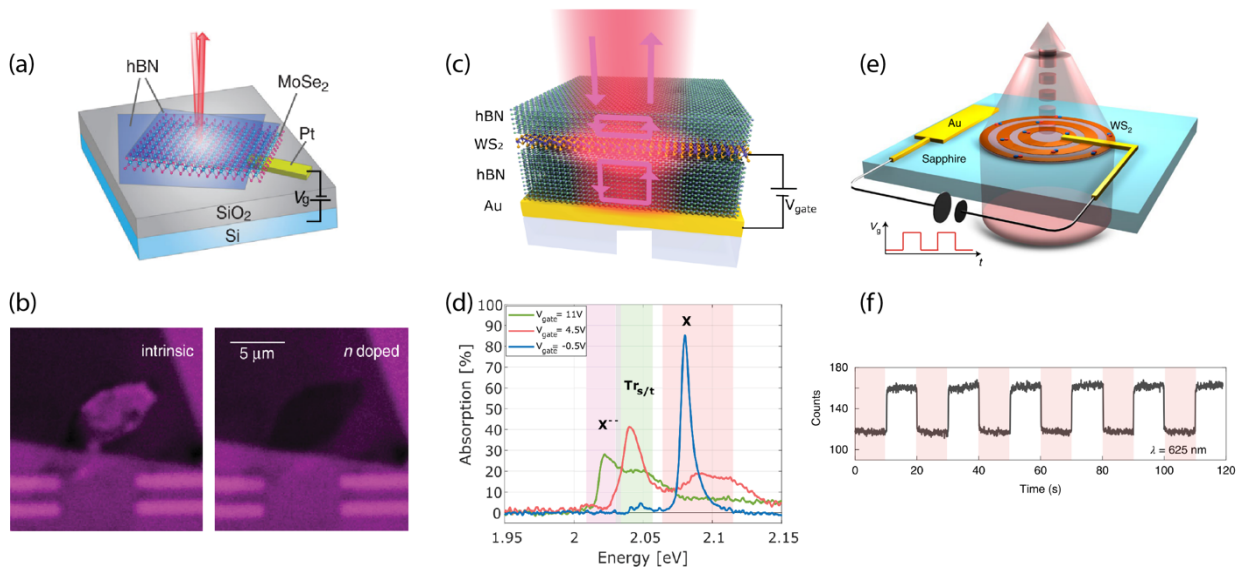

*Figure 8: Exciton resonance tuning in nanophotonic devices and metasurfaces.* (a) A monolayer flake $MoSe_2$ encapsulated between two layers of hBN on an oxidized wafer functions as an electrically-switchable mirror. At cryogenic temperatures (4 K) the exciton reflection interferes with the bare cavity reflection and yields an 80% reflection. (b) By injecting electrons into the $MoSe_2$, the reflectivity can be suppressed. Images reproduced from ref.[203] (c) An encapsulated monolayer $WS_2$ flake in an engineered optical cavity shows electrically-tunable and near-unity absorption (d). By controlling the carrier density, the exciton's non-radiative decay rate can be tuned to match the radiative rate to achieve critical coupling. Images reproduced from ref.[204] (e) A large-area monolayer $WS_2$ is patterned into an atomically-thin metasurface zone plate lens. (f) By switching

*the gate voltage on (red shaded, $V_g$=3 V) and off (white shaded, $V_g$=0 V) the intensity in the focus is modulated by 33% at the exciton wavelength. Images reproduced from ref.[230]*

Atomically-thin tunable absorber

The electrical control over the exciton resonance can also be leveraged to achieve tunable and near-unity absorption in a single monolayer. Epstein et al.[204] demonstrate how a high-quality monolayer flake of WS$_2$ in an optimized photonic environment can achieve near-unity absorption by accurate controlling the balance between radiative and non-radiative decay rates of the exciton state (Fig. 8c). The quantum efficiency of the exciton is highly sensitive to the concentration of free carriers in the 2D semiconductor, since excitons can react with a free carrier to form a charged exciton (called a *trion*)[48]. Trions primarily decay non-radiatively and thus effectively function as a non-radiative decay channel for the neutral exciton state.

Through careful tuning of the background density of free carriers, the quantum efficiency of the exciton can be controlled at will[231]. Using this, Epstein et al. carefully control the relative decay rate to match the non-radiative decay rate to reach the critical coupling condition. Since this balance is easily perturbed by changing the free carrier concentration, very large changes in the absorption can be achieved through external gating (Fig. 8d). Similar to the electrically-switchable mirrors, the tunable absorber is a nanophotonic device formed from microscale flakes operating under cryogenic conditions.

Atomically-thin tunable zone plate lens

The nanophotonic devices discussed so far are all based on high-quality exfoliated and encapsulated flakes at cryogenic temperatures where phonon interactions are suppressed. In the context of metasurfaces however, these first demonstrations are inspiring but not scalable to large areas and ambient conditions. At the same time, the exciton resonances are stable even at room temperature, albeit with smaller oscillator strengths. This prompts the question whether excitons could also be leveraged in large area metasurface optical elements to tune their room-temperature optical response?

In a recent work[230], the first electrically-tunable large-area monolayer metasurface optical element has been demonstrated (Fig. 8e), that operates at ambient conditions. Large-area monolayer WS$_2$ on sapphire (CVD-grown) is patterned into 1-mm-diameter zone plate lens. The lens design is optimized for wavelengths near the exciton resonance with a 2 mm focal length. Despite the relatively low quality of the CVD-grown WS$_2$, a strong exciton resonance is observed in the optical properties of the material. Using an electrochemical cell, the electron density in the WS$_2$ can be tuned effectively using relatively small gate voltages (~3 V). The gold reference electrode is positioned adjacent to the lens, and both are covered with an ionic liquid. The ionic liquid very effectively screens the electrostatic charges in the material and reference electrode, such that high carrier densities can be achieved. Using this, the exciton resonance can be fully suppressed in a reversible manner.

The monolayer lens forms a bright focus 2 mm above the surface, despite the monolayer thickness of only 0.6 nm. By switching the gate voltage on and off, the intensity in the focus of the lens can be actively modulated by 33% because of the exciton resonance tuning (Fig. 8f). At the same time, the lens is highly transparent at all other wavelengths, providing transparent optical elements. Although the optical efficiency of the lens is relatively small (<1%), this first demonstration of exciton resonance tuning in atomically-thin optical elements highlights the opportunities that excitons offer for tunable flat optics.

## 5. Conclusions and outlook

In summary, we have detailed in the above tutorial that excitons in atomically-thin systems are fundamentally novel in their optical character and present a new opportunity in metasurface optics. Particularly, their strong confinement and oscillator strength in a 2D or layered crystal make them amenable to tuning/control via electric fields, carrier injection, changing dielectric environment, temperature, and strain as well as coupling with other quasiparticles in adjacent media such as plasmons in noble metals. These tuning knobs have been explored to some extent in realizing tunable metasurfaces for phase control and amplitude control. However, in much of these demonstrations the use of exciton as an intrinsic resonance of the electronic structure of the lattice and therefore its ability to couple resonantly to photons to form hybrid exciton-polariton light-matter states has been ignored.

This resonant coupling between excitons and cavity photons has been well exploited in passive metasurfaces that use plasmons or dielectric grating modes or Mie resonances to trap the photons and have often employed multilayer rather than monolayer 2D excitonic semiconductors. While active tuning excitons in multilayer 2D semiconductors is difficult, this presents an opportunity to do further novel optical design to maximize light trapping in monolayer structures such as building multi-quantum well or superlattice structures[217]. Simultaneously, clever optical design can also be employed to use the high tunability of exciton resonances yet avoid the intrinsic loss induced by the strong excitonic absorption in reflective metasurfaces. Finally, coherent exciton scattering in 2D semiconductors presents a nearly pristine opportunity to engineer a new generation of tunable atomically-thin flat optical elements with applications far beyond what has been achieved in reflective and transmissive meta-optics thus far. Despite the relatively low optical efficiency of monolayer optical elements, their high transparency combined with tunable optical function makes them very promising for novel applications in beam tapping, augmented reality, and free-space optical communication[232].

While the above opportunities for using excitons in 2D semiconductors for novel meta-optical elements are indeed promising, future applications of excitons on nanophotonic and optoelectronic devices require scalable materials and reliable devices. In this regard, growth and engineering of 2D semiconductors with strict control over crystalline quality, stoichiometry and thickness over wafer scales is one of the most critical challenges. This is primarily because the optical constants and hence exciton resonances in 2D semiconductors are highly sensitive to these three intrinsic parameters related to the material structure. Monolayer 2D semiconductors were first isolated in 2005[37] and have been investigated in detail since 2010[40,233]. Despite this, their wafer-scale growth with controlled thickness and near single crystalline character has become possible for a limited number of them only recently[234–236]. Therefore, significant research is required on this front to render them into a scalable optical materials platform with control over doping, lateral and vertical heterostructure formation. Nevertheless, the opportunities for metasurfaces at the horizon are immense given the number of excitonic 2D materials and the diversity in their optical behavior, resonant energies, and exciton properties.

**Acknowledgements**

L.G. and J.v.d.G are supported by the Dutch Research Council (NWO) and by the NWO VIDI grant (project number VI.Vidi.203.027). J.L. and D.J. acknowledge partial support from the US Army Research Office (ARO) under contract number W911NF-19-1-0109, the Air Force Office of Scientific Research (AFOSR) grant number FA9550-21-1-0035 and also from the Asian Office of Aerospace Research and Development of the AFOSR with grant numbers FA2386-20-1-4074 and FA2386-21-1-4063. J. L. is also supported by the Ashton Fellowship administered by the School of Engineering and Applied Sciences of the University of Pennsylvania.

**Conflict of interest**

The authors have no conflicts to disclose.

**Data availability**

Data sharing is not applicable to this article as no new data were created or analyzed in this study.